%% file: paper.tex
\documentclass[conference,UKenglish]{IEEEtran}
\IEEEoverridecommandlockouts%

\usepackage{babel}
\usepackage{cite}
\usepackage{amsmath,amssymb,amsfonts}
\usepackage{algorithmic}
\usepackage{graphicx}
\usepackage{textcomp}
\usepackage{xcolor}
\usepackage{todonotes}
\usepackage{csquotes}
\usepackage[
add-integer-zero=false,%
free-standing-units=true,%
list-final-separator={, and },%
round-mode=figures,%
round-precision=3,%
separate-uncertainty=true,%
uncertainty-mode=separate,%
mode=match,%
]{siunitx}
\usepackage{microtype}
\usepackage{url}
\usepackage{subcaption}
\usepackage{listings}
\usepackage{hyperref}
\usepackage[capitalise]{cleveref}
\usepackage{multirow}
\usepackage{booktabs}
\usepackage{balance}

\def\BibTeX{{\rm B\kern-.05em{\sc i\kern-.025em b}\kern-.08em
		T\kern-.1667em\lower.7ex\hbox{E}\kern-.125emX}}

\newcommand{\exnum}[1]{\num[round-mode=none]{#1}}
\newcommand{\perc}[1]{\qty[round-mode=none]{#1}{\percent}}
\newcommand{\dollar}[1]{US-\$\,\exnum{#1}}

\lstdefinelanguage{diff}{
	basicstyle=\ttfamily\small,
	morecomment=[f][\color{green}]{+\ },
	morecomment=[f][\color{red}]{-\ }
}

\newcommand{\summary}[2]{%
	\vspace{-0.2cm}%
	\begin{center}%
		\colorbox{gray!20}{%
			\parbox{\linewidth}{%
				\textbf{\textsf{Summary (\textit{#1})}:}~%
				#2%
			}%
		}%
	\end{center}%
}

\newcommand{\summarybottom}[2]{%
	\noindent
	\setlength{\fboxsep}{2pt} %
	\colorbox{gray!20}{%
		\parbox{\dimexpr\linewidth-2\fboxsep}{%
			\textbf{\textsf{Summary (\textit{#1})}:}~%
			#2%
		}%
	}%
	\setlength{\fboxsep}{3pt} %
}

\begin{document}
	
	\title{Mutation Testing via Iterative Large Language Model-Driven Scientific Debugging
	}
	
	\author{\IEEEauthorblockN{Philipp Straubinger}
		\IEEEauthorblockA{\textit{University of Passau} \\
			Passau, Germany}
		\and
		\IEEEauthorblockN{Marvin Kreis}
		\IEEEauthorblockA{\textit{University of Passau} \\
			Passau, Germany}
		\and
		\IEEEauthorblockN{Stephan Lukasczyk\textsuperscript{*}}
		\IEEEauthorblockA{\textit{JetBrains Research} \\
			Munich, Germany}
		\and
		\IEEEauthorblockN{Gordon Fraser}
		\IEEEauthorblockA{\textit{University of Passau} \\
			Passau, Germany}
	 \thanks{\null\textsuperscript{*} Research partially carried out while with the University of Passau.}
	}
	
	\maketitle
	
	\begin{abstract}

          Large Language Models (LLMs) can generate plausible test
          code. Intuitively they generate this by imitating tests seen
          in their training data, rather than reasoning about
          execution semantics. However, such reasoning is important
          when applying mutation testing, where individual tests need
          to demonstrate differences in program behavior between a
          program and specific artificial defects (mutants).
          In this paper, we evaluate whether \textit{Scientific
            Debugging}, which has been shown to help LLMs when
          debugging, can also help them to generate tests for
          mutants. In the resulting approach, LLMs form hypotheses
          about how to kill specific mutants, and then iteratively
          generate and refine tests until they succeed, all with
          detailed explanations for each step.
          We compare this method to three baselines: (1) directly
          asking the LLM to generate tests, (2) repeatedly querying
          the LLM when tests fail, and (3) search-based test
          generation with Pynguin. Our experiments evaluate these
          methods based on several factors, including mutation score,
          code coverage, success rate, %
          and the ability to identify equivalent mutants. The results
          demonstrate that LLMs, although requiring higher
          computational cost, consistently outperform Pynguin in
          generating tests with better fault detection and
          coverage. Importantly, we observe that the iterative
          refinement of test cases is important for achieving
          high-quality test suites.
		
	\end{abstract}
	
	
	\section{Introduction}
	
	\input{sections/introduction}

	\section{Background}
	
	\input{sections/background}

	\section{Scientific Test Generation}\label{sec:scientific}
	
	\input{sections/generation}

	\section{Evaluation}
	
	\input{sections/evaluation}

	\section{Discussion}
	
	\input{sections/discussion}

	\section{Related Work}
	
	\input{sections/related}

	\section{Conclusions}
	
	\input{sections/conclusions}

	In order to support experiment replications and further research, our dataset is available at:
	
	\begin{center}
		\url{https://www.doi.org/10.6084/m9.figshare.28550855}
	\end{center}
	
	\balance
	
	\bibliographystyle{ieeetr}
	\bibliography{bib}
	
\end{document}

%% file: sections/introduction.tex
Ensuring the quality of software is a cornerstone of modern software
development, as bugs and vulnerabilities can lead to significant
financial, operational, and reputational
damages~\cite{pooreport,DBLP:journals/informatics/PereraJMO22,DBLP:journals/tse/TelangW07}. As
manual creation of tests is often considered labor-intensive,
time-consuming, and error-prone, automated test generation has emerged
as a promising
solution~\cite{DBLP:conf/qrs/KlammerR17,DBLP:conf/sbes/SouzaM20}.
Although traditional methods for automated test generation, such as
search-based methods or symbolic execution, have proven their utility,
they still face limitations in handling complex code structures,
achieving high coverage, and adapting to diverse software
ecosystems~\cite{DBLP:journals/cacm/CadarS13,DBLP:conf/osdi/CadarDE08}.
Large language models~(LLMs), such as OpenAI's ChatGPT series, and
their ability to perform complex tasks by understanding and generating
human-like
text~\cite{DBLP:conf/naacl/ZhengZZXGXC24,DBLP:journals/frai/OrruPCG23},
offer the potential to overcome these limitations.

LLMs can generate plausible test code, which intuitively they do by
imitating tests seen in their training data. While this results in
readable and realistic tests, it remains unclear whether this approach
is best suited for generating test cases targeting specific scenarios
or faults. In particular, mutation testing is a common strategy used
to ensure that test suites are produced that can distinguish a
program from as many as possible artificial defects, i.e., mutants.
To do so effectively, LLMs would need to reason about
execution semantics and how they are affected by individual
bugs. Interestingly, recent work has demonstrated a promising approach
to reason about bugs with LLMs, by simulating the scientific method of
debugging~\cite{DBLP:journals/corr/abs-2304-02195} (Scientific
Debugging). Consequently, this raises the question of whether this
approach can also achieve more effective mutation testing
using LLMs.

To shed light on this question, in this paper, we introduce an
LLM-based test generation approach for Python using the concept of
Scientific Debugging. The process iterates through hypothesis
formulation, experimentation, and conclusion cycles to generate test
cases that identify behavioral discrepancies between a program under
test~(PUT) and a mutant. The workflow is highly automated and
consistent, with strict formatting requirements to ensure accurate
parsing and execution, and clear documentation that supports
reproducibility. The methodology emphasizes efficient test case
design, where successful test cases highlight mutant behavior by
failing on the mutant while passing on the PUT\@. If extensive testing
does not detect any differences, the approach considers the mutant to
be functionally equivalent to the PUT\@.

In detail, our main contributions are as follows:
\begin{itemize}
	\item We introduce a prompt framework for generating tests based on
	scientific debugging principles.
	\item We empirically evaluate LLM-based test generation on a carefully curated dataset of Python projects, leveraging mutation analysis to evaluate test quality and the practical utility of LLMs in real-world scenarios.
	\item We compare our results with Pynguin, a state-of-the-art search-based test-generation tool for Python~\cite{DBLP:conf/icse/LukasczykF22}.
	\item We identify critical factors that influence the performance of LLMs in automated testing, such as prompt design, and provide detailed analyses to highlight their applicability and limitations.
\end{itemize}

The results show that LLM-based approaches outperform Pynguin in both branch coverage and mutation score, with similar results in line coverage. Iterative prompting is effective in generating comprehensive test suites, yielding consistent results regardless of the refinement strategy. The baseline approach, only prompting the LLM once, is more cost-effective, while scientific approaches are more expensive due to higher token usage. Despite the higher cost, LLM-based methods with iterations maintain similar success rates. Pynguin is cheaper to run but the generated tests achieve lower coverage and mutation scores. Additionally, scientific approaches tend to label more mutants as equivalent if no test is found for them, but most of these mutants are not actually equivalent.

%% file: sections/background.tex
\subsection{Large Language Models}

Large Language Models~(LLMs) have significantly advanced Natural Language Processing~(NLP) and Programming Language Processing~(PLP)~\cite{DBLP:conf/nips/VaswaniSPUJGKP17}. Built on the Transformer architecture~\cite{DBLP:conf/nips/VaswaniSPUJGKP17}, these models are pre-trained on massive text corpora~\cite{DBLP:journals/tse/MillerS76,DBLP:conf/nips/Ouyang0JAWMZASR22} using self-supervised techniques~\cite{DBLP:conf/naacl/AhmadCRC21,DBLP:conf/sigsoft/ChakrabortyADDR22}. This enables LLMs to perform NLP tasks such as machine translation, text generation, and question answering~\cite{DBLP:conf/nips/BrownMRSKDNSSAA20,DBLP:conf/nips/Ouyang0JAWMZASR22}.
In PLP, example applications of LLMs include code completion~\cite{DBLP:journals/tse/CiniselliCPMAPP22}, program repair~\cite{DBLP:conf/kbse/LiLFMXCL22}, and code generation~\cite{DBLP:conf/sigsoft/SvyatkovskiyDFS20}. Despite their success, they face challenges such as low pass rates and errors during code generation, largely due to their reliance on pre-training datasets~\cite{DBLP:journals/pacmse/Yuan0DW00L24,DBLP:journals/tse/SchaferNET24,DBLP:journals/corr/abs-2406-18181,DBLP:journals/corr/abs-2412-14137}.

\subsection{Mutation Testing}

Mutation testing is a fault-based software testing technique designed to assess the quality and effectiveness of test suites by introducing artificial faults, known as mutants, into a program~\cite{DBLP:journals/tse/JiaH11}. These mutants are slight variations of the original code, often created by applying small syntactic changes, such as modifying operators or variable types. The goal is to determine if the existing test cases can detect these artificial faults, thereby estimating the ability of the test suite to identify real-world defects~\cite{DBLP:journals/tse/JiaH11}. If a test case fails when executed against a mutant, it \textit{kills} the mutant; conversely, if no test cases detect a mutant, it \textit{survives}, indicating potential weaknesses in the test suite.

\subsection{Scientific Debugging}

The scientific method offers a structured approach to
debugging~\cite{zeller2009programs}. It starts with generating
hypotheses that can explain the root cause of a bug based on known
observations. Developers then predict expected outcomes if the
hypothesis is correct, design experiments to test these predictions,
and observe the actual results from those experiments. The outcome
either confirms or rejects this hypothesis. This cycle repeats until
the bug is identified and fixed.

Recent work on \emph{Scientific
	Debugging}~\cite{DBLP:journals/corr/abs-2304-02195} demonstrated
that this approach can be automated using LLMs: LLMs generate
hypotheses based on the buggy code and error messages, design and
suggest experiments such as specific debugger commands or code
modifications to validate these hypotheses and analyze the
results. Through their ability to process large amounts of information
and emulate reasoning patterns, LLMs can accelerate the debugging
process by iteratively refining their understanding of the bug, just
as a developer would. Furthermore, LLMs generate explanations for each
step, providing clear reasoning for why certain fixes are attempted,
thus making the debugging process more transparent and easier to
follow for developers. This integration of LLMs into the Scientific
Debugging process not only automates the identification of bugs but
also improves the explainability and efficiency of automated debugging
tools.

\subsection{Search-based Test Generation}

Search-based testing has emerged as an effective test generation
approach applicable to many different
domains~\cite{DBLP:journals/csur/HarmanMZ12}, including the generation
of unit tests for Python, our aim in this paper.  In search-based
testing, evolutionary algorithms maintain a population of candidate
solutions and iteratively refine them using crossover, mutation, and
selection~\cite{DBLP:series/ncs/EibenS15} following the principles of
Darwinian evolution~\cite{Dar59} known from nature. For test
generation, individuals are test cases or test suites, and the
evolution is guided by fitness functions based on code
coverage~\cite{DBLP:journals/stvr/McMinn04}. Several tools, e.g.,
EvoSuite~\cite{DBLP:conf/sigsoft/FraserA11} for Java or
Pynguin~\cite{DBLP:conf/icse/LukasczykF22} for Python exist and
provide implementations of these concepts for automated test
generation.

%% file: sections/generation.tex
\subsection{Scientific Debugging Process}

Our methodology applies a systematic approach to debugging Python
programs. First, we present the LLM with the targeted mutant, as well
as its containing module, then ask it to iteratively generate
hypotheses and experiments until the mutant behavior can be
explained. Hypotheses describe the expected behavior of the program or
the mutant, while experiments consist of code that validates these
hypotheses by running tests. We execute these experiments and
inform the LLM about the outcome. Then, we prompt the LLM again to perform
the next iteration if the mutant has not been detected.
Otherwise, the LLM is instructed to generate a test that exploits
the knowledge gained from prior experiments. If the test is not
able to detect the mutant, the LLM can choose to refine the test or
continue iterating with experiments.

This debugging workflow is designed for automation and consistency,
with strict output formatting requirements to ensure accurate parsing
and execution. Each stage of the process is explicitly documented
using a predefined markdown-based structure, which organizes
information under clear, hierarchical headings. This structure
supports the systematic nature of scientific debugging, ensuring
clarity and reproducibility throughout the debugging process.

The following steps are part of scientific debugging\@:

\paragraph{Initial Prompting}

We present the LLM with an explanation of the test generation process, as well as the targeted \emph{mutant} and its containing module, referred to as the \emph{Program under Test~(PUT)}.

\paragraph{Hypothesis Formulation}

Each hypothesis describes an assumption or observation regarding the behavior of the mutant, informed by prior experiments or an initial analysis of the code. A hypothesis should be specific and predictive; for example, \enquote{For input [x], the baseline is expected to produce [result], whereas the mutant will yield [a different result].} These predictions serve as a foundation for subsequent experiments and are iteratively refined based on the outcomes of these experiments. A hypothesis should, if accepted, help the LLM reason about the mutant.

\paragraph{Experimentation}

Experiments consist of Python functions in which the LLM evaluates the specific behavior of the target code. They typically call a target function, print some results to the console, and perform assertions on those results based on the current hypothesis. We execute each experiment independently on both the PUT and the mutant, and compare the resulting outputs to identify discrepancies; such discrepancies can be differences in returned values, raised exceptions, or execution timeouts. This requires carefully crafted experiments that investigate particular conditions influenced by the mutant.

\paragraph{Conclusions}

After each experiment, we relay the execution results back to the LLM and ask it to write a short conclusion. The LLM can then either choose to continue formulating new hypotheses and experiments or generate a test candidate. If we encounter an experiment that does not compile, we provide the LLM with the compile error and ask it to try again.

\paragraph{Test Cases}

Once a behavioral discrepancy between the PUT and the mutant is highlighted by an experiment, a test case can be made from the experiment by removing unnecessary code and adding assertions as well as comments. We do not strictly require tests to follow a successful experiment, but we do instruct the LLM to follow this suggestion. A successful test case needs to exploit the found behavioral discrepancy with assertion errors, uncaught exceptions, or timeouts when executed on the mutant that do not occur when executed on the PUT\@. As a result, such a test case cannot directly replicate the mutant's implementation nor rely on hardcoded assumptions on the program itself.

The development of a test that detects the mutant ends the iterative process. If the LLM submits an ineffective test, we give it execution results, similarly to experiments. The LLM can choose whether to continue with hypotheses and experiments or with another test candidate.

\paragraph{Handling Equivalent Mutants}

At any point, the LLM can choose to claim the mutant as equivalent. In such cases, a detailed explanation is provided, including reasoning and evidence from the experiments, to substantiate the equivalence claim. For our evaluation, we ask the LLM to continue the test generation process at this point such that we can assess the ability to kill mutants and claim equivalences separately. Under regular circumstances, this would conclude the conversation.

\paragraph{Key Design Principles}

Each hypothesis provides a prediction of the mutant's behavior, which is rigorously tested through controlled experiments. We design these experiments and tests to be compatible with automated systems, which parse and execute the code. We follow strict output-formatting guidelines to achieve this, e.g., consistent use of Markdown headings or well-structured code snippets. Each experiment targets a specific behavioral aspect of the mutant: it minimizes external dependencies and enhances clarity. We keep experiments concise by typically testing one or two inputs at a time, thus ensuring clarity and reducing the potential for errors.

The methodology avoids recreating the mutant or the PUT, focusing solely on detecting behavioral differences. We document each step of the debugging process thoroughly, including the rationale behind hypotheses, the design of experiments, and the interpretation of results. This documentation ensures the transparency of the process and facilitates the reproducibility and extension of findings in future research.

We require tests to complete within a \qty[round-mode=none]{5}{\second} execution window. Timeouts, errors, and exceptions can be significant indicators of mutant behavior. We limit the number of iterations to ten to prevent excessively long runs; this ensures that the debugging process remains efficient, and the LLM can find a solution within a reasonable time.

\subsection{Prompt Design}

\Cref{lst:prompt}~shows a shortened version of the scientific debugging explanation that we include in our main prompt. The full prompt also includes a short explanation about the automated iterative process and the expected output format, as well as a few pitfalls we want the LLM to avoid (like assuming the mutant and the regular PUT are simultaneously available for import). We use additional prompt templates to react to LLM-generated messages. This includes replying with execution results to each experiment/test and responding to invalid messages and non-compilable code.

When using iterative prompting, the entire conversation history has to be digested by the LLM for every new message it generates. Therefore, we avoid unnecessary LLM completions by bundling messages together. One generated message will typically include a conclusion, a hypothesis, and an experiment.

\begin{lstlisting}[float, language={}, caption={Prompt excerpt with instructions for the scientific debugging method}, captionpos=b, label=lst:prompt, linewidth=\columnwidth, basicstyle=\ttfamily\footnotesize, escapeinside=||, frame=single, aboveskip=10pt]
# Scientific Debugging
Scientific debugging is a systematic debugging ap-
proach based on the scientific method. The process
follows a loop of:
- Hypothesis
- Experiment
- Conclusion

## Hypotheses
Each hypothesis should describe an assumption you
have about the code. |\textnormal{\ldots}|
Predict exactly what will happen. Avoid broad pre-
dictions like "Under any of the given inputs, the
mutant will behave differently". |\textnormal{\ldots}|

## Experiments
After stating a hypothesis, you create an experi-
ment to test it. Each experiment will contain a
Python test case, which imports and calls the tar-
get code. |\textnormal{\ldots}|
Each experiment should contain a relevant predic-
tion based on your hypothesis and a way to verify
that prediction based on the output. |\textnormal{\ldots}|
### Example Experiment |\textnormal{\ldots}|

## Conclusions
After every experiment, write a conclusion that
summarizes the results. Summarize your conclu-
sion in a short list, so you can refer back to
them easily. |\textnormal{\ldots}|

## Tests
Once you have found any inputs that cause a diff-
erence in behavior, you can write a test that
kills the mutant. |\textnormal{\ldots}|
The test kills the mutant if, and only if, the
test passes when executed with the **Baseline**
and fails when executed with the **Mutant**. |\textnormal{\ldots}|
Include a relevant docstring comment with a sum-
mary of your findings. |\textnormal{\ldots}|
### Example Test |\textnormal{\ldots}|

## Equivalent Mutants
Some mutants may be equivalent. Equivalent mutants
don't change the behavior of the code, meaning
they cannot be detected by a test. |\textnormal{\ldots}|
If you believe a mutant to be equivalent, write
the `## Equivalent Mutant` headline and give a
short description of why you think the mutant is
equivalent. |\textnormal{\ldots}|
Example: |\textnormal{\ldots}|
\end{lstlisting}

%% file: sections/evaluation.tex
To evaluate LLM-based mutation testing using the proposed methodology,
we ask the following research questions:

\begin{itemize}
	\item RQ1: How does LLM-based test generation perform compared to Pynguin?
	\item RQ2: How efficient is LLM-based test generation?
	\item RQ3: Can LLM-based test generation reliably detect equivalent mutants?
\end{itemize}

\subsection{Experiment Setup}\label{sec:setup}

To address our research questions, we use variants of the LLM-based approach via the OpenAI API with version \texttt{gpt-4o-mini-2024-07-18}\@:

\begin{itemize}
	\item Baseline Approach: We directly ask the LLM to generate a test that can eliminate the specified mutant mentioned in the prompt.
	\item Iterative Approach: Here, we ask the LLM to generate a test for a given mutant and then execute the test on both the subject under test~(SUT) and the mutant. If the test does not produce the expected results (i.e., it should pass for the SUT and fail for the mutant), we request the LLM to refine the test. We repeat this process to at most ten iterations until either the test performs as expected or we reach the iteration limit.
	\item Scientific (0-shot) Approach: We apply the scientific test generation method as described in \cref{sec:scientific} with an iteration limit of 10.
	\item Scientific (1-shot) Approach: We use the same prompts and configurations as in the Scientific (0-shot) approach but with an additional example of a test generation loop.
\end{itemize}

\begin{table*}[t]
	\caption{Overview of the projects used in this paper}
	\label{tab:projects}
	\resizebox{\textwidth}{!}{
		\begin{tabular}{@{} l rrrr SSSSSS @{}}
			\toprule
			\multicolumn{1}{l}{\multirow{2}{*}{Project}} & \multicolumn{1}{c}{\multirow{2}{*}{Modules}} & \multicolumn{1}{c}{\multirow{2}{*}{LoC}} & \multicolumn{1}{c}{\multirow{2}{*}{Total mutants}} & \multicolumn{1}{c}{\multirow{2}{*}{Sampled mutants}} & \multicolumn{6}{c}{Costs (US-\$)}                                                                                                                                                                                                \\ \cmidrule(l{2pt}r{2pt}){6-11}
			\multicolumn{1}{c}{}                         & \multicolumn{1}{c}{}                         & \multicolumn{1}{c}{}                     & \multicolumn{1}{c}{}                               & \multicolumn{1}{c}{}                                 & \multicolumn{1}{c}{Baseline} & \multicolumn{1}{c}{Iterative} & \multicolumn{1}{c}{Scientific (0-shot)} & \multicolumn{1}{c}{Scientific (1-shot)} & \multicolumn{1}{c}{Pynguin (avg)} & \multicolumn{1}{c}{Pynguin (sum)} \\ \midrule
			apimd                                        & 3                                            & 446                                      & 690                                                & 690                                                  & 0.612956                     & 3.86783                       & 4.616916                                & 5.24893                                 & 0.018057                          & 0.523643                          \\
			codetiming                                   & 3                                            & 91                                       & 74                                                 & 74                                                   & 0.023523                     & 0.195641                      & 0.353698                                & 0.378896                                & 0.008107                          & 0.243214                          \\
			dataclasses-json                             & 7                                            & 1,009                                     & 655                                                & 655                                                  & 0.388398                     & 3.114447                      & 4.429182                                & 5.082354                                & 0.026748                          & 0.802452                          \\
			docstring\_parser                            & 11                                           & 1,665                                     & 3,829                                               & 569                                                  & 0.292733                     & 1.820551                      & 2.509935                                & 2.798515                                & 0.004418                          & 0.132534                          \\
			flake8                                       & 26                                           & 3,551                                     & 2,722                                               & 999                                                  & 0.523106                     & 5.401514                      & 7.717868                                & 8.678167                                & 0.130215                          & 3.776247                          \\
			flutes                                       & 13                                           & 1,501                                     & 1,949                                               & 1,000                                                 & 1.293902                     & 7.606816                      & 9.671575                                & 10.182957                               & 0.039297                          & 1.178903                          \\
			flutils                                      & 17                                           & 2,099                                     & 2,213                                               & 1,000                                                 & 0.456861                     & 4.519201                      & 6.634037                                & 7.381829                                & 0.065117                          & 1.693041                          \\
			httpie                                       & 32                                           & 3,176                                     & 2,345                                               & 1,000                                                 & 0.394936                     & 4.562005                      & 6.276648                                & 7.305439                                & 0.115317                          & 3.459505                          \\
			pdir2                                        & 8                                            & 613                                      & 393                                                & 393                                                  & 0.152105                     & 1.421908                      & 1.935523                                & 2.596913                                &   {--}                                & {--}                                  \\
			python-string-utils                          & 6                                            & 581                                      & 1,122                                               & 1,000                                                 & 0.952624                     & 4.54425                       & 8.240349                                & 8.935906                                & 0.021703                          & 0.651083                          \\ \midrule
            Total                                        & 126                                          & 14,732                                  & 15,992                                            & 7,380                                                 & 5.091144                     & 37.054163                     & 52.385731                               & 58.589906                               & 0.428979                          & 12.460622                         \\ \bottomrule
		\end{tabular}
	}
\end{table*}

We selected a subset of ten Python projects from a larger dataset of 20 projects previously used in a study with Pynguin~\cite{DBLP:journals/ese/LukasczykKF23}. By choosing from prior work on Pynguin the projects are more likely to be compatible with Pynguin, and we also limited the sample to modules with 1,000 lines of code or fewer, avoiding input sizes that exceed API limits. Despite this selection process, for the \texttt{pdir2} project, Pynguin did not yield any results: after Pynguin finishes the test generation, it attempts to reload the module to compute the final results, e.g., coverage, for the generated test suite. The \texttt{pdir2} project's implementation overrides internal data structures of Python, which leads to a crash of the Python interpreter during the aforementioned reload attempt. Thus, Pynguin is neither able to dump test cases nor coverage results for this project.

We used Cosmic Ray\footnote{\url{https://cosmic-ray.readthedocs.io/en/latest/}} to generate mutants for all modules across these ten projects. For test generation as well as evaluation, we then randomly selected up to 1,000 mutants from each project (see \cref{tab:projects}). To eliminate flaky tests, we execute each generated test suite on the unmodified PUT one hundred times, excluding any failing tests in the process. All test executions, both during test generation and during evaluation, are performed inside of containers, to reduce side effects, and to prevent potentially harmful tests from interacting with the host system. During test generation, each test is executed on a fresh copy of the PUT.

To assess the significance of the differences, we used the exact Wilcoxon-Mann-Whitney test~\cite{10.1214/aoms/1177730491} with \(\alpha=0.05\). We report measured results rounded to three significant digits~\cite{DBLP:journals/sttt/BeyerLW19}.

\subsubsection{RQ1: How does LLM-based test generation perform compared to Pynguin}

To answer this question, we apply our four LLM-based approaches alongside Pynguin~\cite{DBLP:conf/icse/LukasczykF22} to all sampled mutants. We applied Pynguin revision \texttt{fd9a6e96} from its GitHub repository\footnote{\url{https://github.com/se2p/pynguin}} in its default settings on each module in the evaluation projects, i.e., with the DynaMOSA~\cite{DBLP:journals/tse/PanichellaKT18} test-generation algorithm and a timeout of \qty{600}{\second}. To isolate the processes, we execute Pynguin inside Podman containers and run the experiments on dedicated compute servers equipped with two Intel Xeon E5-2620v4 CPUs per node and \qty{256}{\giga\byte} RAM, running Debian Linux. Each run of Pynguin is bound to one CPU core and \qty[round-mode=none]{4}{\giga\byte} RAM\@. In line with recommendations for the evaluation of randomized algorithms~\cite{DBLP:journals/stvr/ArcuriB14}, we execute Pynguin 30 times on each subject system.  We then compare the results based on (1) branch and (2) line coverage per project, (3)~mutation score calculated on the sampled mutants per project, and (4) number of generated tests per project.%

\subsubsection{RQ2: How efficient is LLM-based test generation?}

For this research question, we compare the efficiency of the four approaches by analyzing (1) the costs, (2) the number of iterations to finish the execution, and (3) the outcome of an execution (success, fail, equivalence claim). For Pynguin, we estimate the costs based on the total execution times of the tool, if we had executed it on an AWS-EC2 machine with similar specifications to what we assigned. We do this because our compute cluster does not provide the necessary information. The machine type for AWS-EC2 is called \texttt{t4g.medium} and costs (at the time of writing) US-\$\,\num[round-precision=4,round-mode=places]{0.0384} per hour. Please note that the execution of Pynguin inside Podman causes an overhead that cannot be quantified easily.%

\subsubsection{RQ3: Can LLM-based test generation reliably detect equivalent mutants?}

To investigate this, we randomly select up to five mutants labeled as equivalent by either Iterative, Scientific (0-shot), or Scientific (1-shot) for each project if available, along with five unkilled mutants per project. We manually review these mutants to verify the accuracy of their equivalence claims. This involves comparing the mutants side by side and creating killing tests when we determine that an equivalence claim is incorrect. This analysis allows us to evaluate the effectiveness of each approach in identifying equivalent mutants. Additionally, we analyze the frequency of equivalence claims across the four approaches.

\subsection{Threats to Validity}

\paragraph{Internal Validity}

Our sample of up to 1,000 mutants per project may suffer from selection bias. To mitigate this, we applied random selection to minimize potential biases. Execution environment variability posed a risk, as variations could influence execution behaviors. To address this, we standardized conditions by using containerized environments~(Podman).

\paragraph{External Validity}

We used a limited dataset of Python projects sampled from existing
studies on Pynguin~\cite{DBLP:journals/ese/LukasczykKF23}, which
restricts the generalisability of our findings to other languages or
larger systems. However, the chosen projects represent a diverse range
of domains and complexities, providing a solid starting
point. Additionally, our analysis of equivalent mutants relies on Python-specific
language features.  Future work shall explore the approach on other
languages to assess consistency across ecosystems.

\paragraph{Construct Validity}

The use of coverage metrics, such as branch and line coverage, may not fully capture the quality of the generated test cases, as they do not directly measure fault detection capabilities. To address this, we supplemented these metrics with mutation scores and the number of equivalent mutants identified, providing a more comprehensive evaluation. LLM prompt design also impacts the quality of test generation, as suboptimal prompts could result in less effective test cases. To mitigate this, we iteratively refined prompts and compared multiple designs to identify effective strategies.

\subsection{Results}

\subsubsection{RQ1: How does LLM-based test generation perform compared to Pynguin?}\label{sec:rq1}

\begin{figure*}[t]
	\centering
	\begin{subfigure}[t]{0.24\textwidth}
		\centering
		\includegraphics[width=\textwidth]{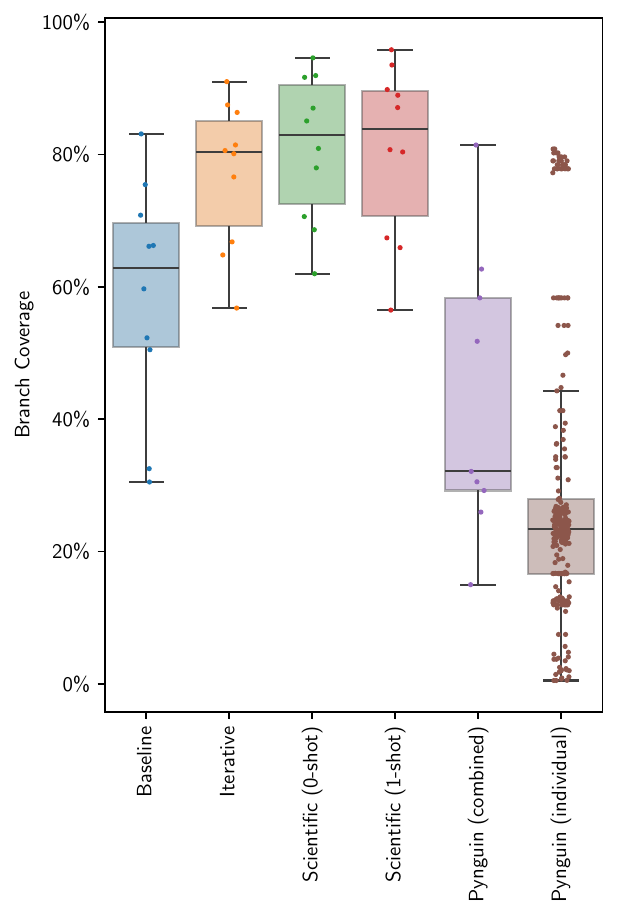}
		\caption{Branch coverage}
		\label{fig:branch}
	\end{subfigure}
	\hfill
	\begin{subfigure}[t]{0.24\textwidth}
		\centering
		\includegraphics[width=\textwidth]{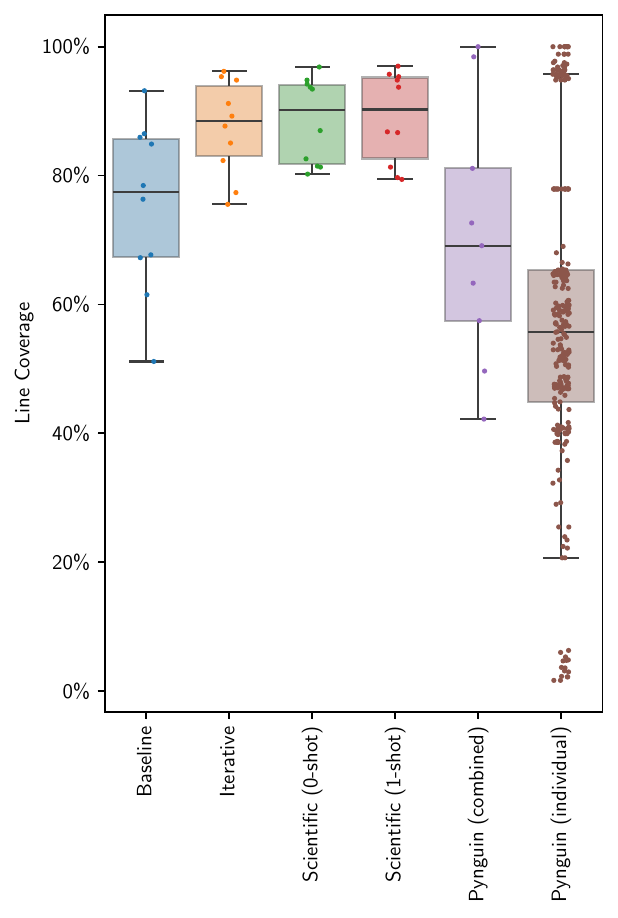}
		\caption{Line coverage}
		\label{fig:line}
	\end{subfigure}
	\hfill
	\begin{subfigure}[t]{0.24\textwidth}
		\centering
		\includegraphics[width=\textwidth]{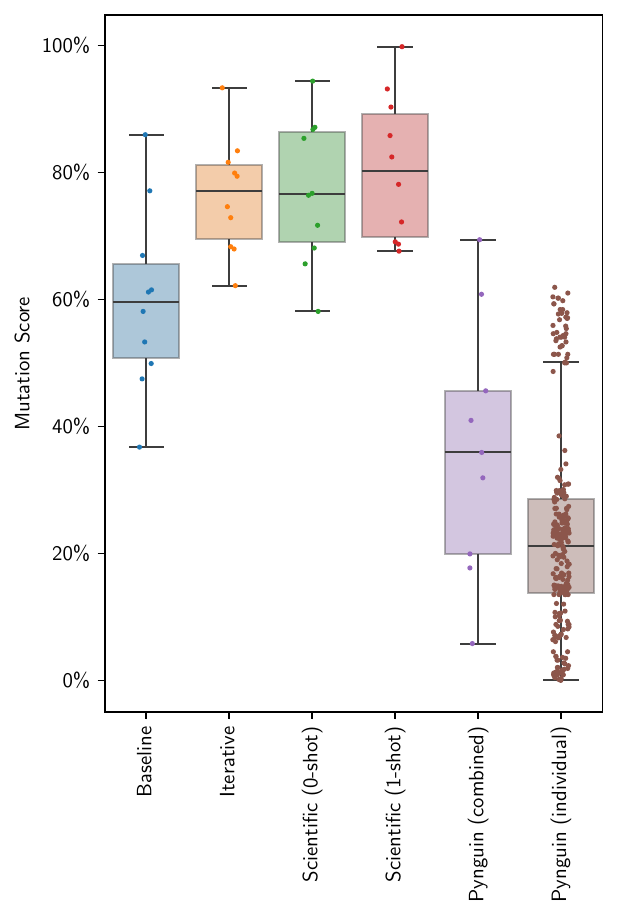}
		\caption{Mutation score}
		\label{fig:mutation}
	\end{subfigure}
	\hfill
	\begin{subfigure}[t]{0.24\textwidth}
		\centering
		\includegraphics[width=\textwidth]{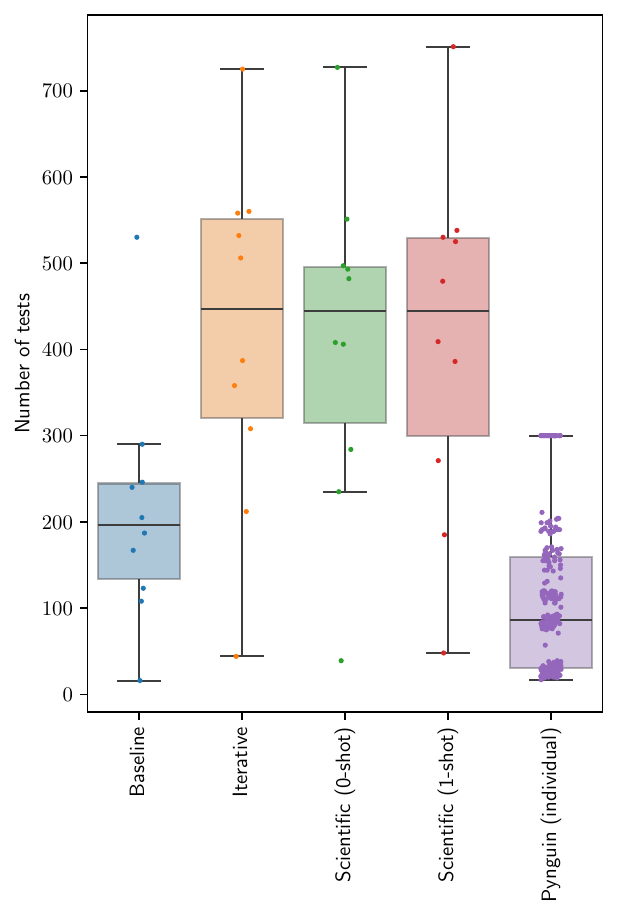}
		\caption{Number of tests}
		\label{fig:tests}
	\end{subfigure}

	\caption{Performance differences, each time along with the individual Pynguin test suites and the combined test suites.}
	\label{fig:rq1}
\end{figure*}

\Cref{fig:rq1} presents the results of the different approaches described in \cref{sec:setup}, including the performance of individual Pynguin runs as well as the combined test suites from all Pynguin runs. \Cref{fig:branch} illustrates the branch coverage achieved by the resulting test suites. Among our approaches, the Baseline achieved the lowest branch coverage, around \perc{55} (\(p<0.05\)). This result is expected since the Baseline does not include refinement iterations. The remaining three approaches achieved nearly identical branch coverage, around \perc{80}. This result is surprising, as the scientific approaches aim to generate tests iteratively based on hypotheses, while the Iterative approach generates new tests for surviving mutants without hypotheses.

A similar pattern emerges when examining line coverage (\cref{fig:line}). The Baseline achieves approximately \perc{75} line coverage, while the other three approaches reach nearly \perc{90}, significantly (\(p<0.05\)) more than the Baseline. The mutation score (\cref{fig:mutation}) follows the same trend, with the Baseline achieving around \perc{60}, and the other approaches reaching about \perc{80}. However, the differences between the Baseline and the other approaches are statistically significant (\(p<0.001\)), which is expected since the approaches are optimized for mutants.

The number of tests generated (\cref{fig:tests}) further supports these observations. The Baseline generates roughly half as many tests as the other three approaches, with this difference being statistically significant. The other approaches each generate about 4,000 tests overall and achieve similar mutation scores. This suggests that the specific way the LLM is prompted matters less, as long as the process is iterative—the results remain comparable.

When comparing the LLM-based approaches with Pynguin, it is evident that the LLM-based methods consistently outperform Pynguin. This holds for both the individual and combined test suites from \exnum{30} runs. The differences in mutation score between the LLM approaches and Pynguin are always statistically significant, although the differences in line coverage are not. Regarding branch coverage, the difference between Pynguin and the Baseline is not significant (\(p = 0.079\)), but it is significant when comparing Pynguin to the other approaches. This is particularly noteworthy because Pynguin focuses on optimizing for branch rather than line coverage. The differences in the number of generated tests are always significant. Based on these findings, we cannot conclude that LLM-based approaches will always outperform search-based approaches like Pynguin. However, there is a clear trend favoring LLM-based methods in terms of branch coverage and mutation score. Future research should investigate whether these trends hold across a broader range of programs, configurations, and LLMs.

\summary{RQ1}{The LLM-based approaches consistently outperform Pynguin in branch coverage and mutation score, with statistically significant differences, although they show comparable results in line coverage. Among the LLM methods, iterative prompting achieves similar results regardless of the specific refinement strategy, highlighting its effectiveness in generating comprehensive test suites.}

\subsubsection{How efficient is LLM-based test generation?}

\begin{figure}[t]
	\centering
	\begin{subfigure}[t]{0.49\columnwidth}
		\centering
		\includegraphics[width=\textwidth]{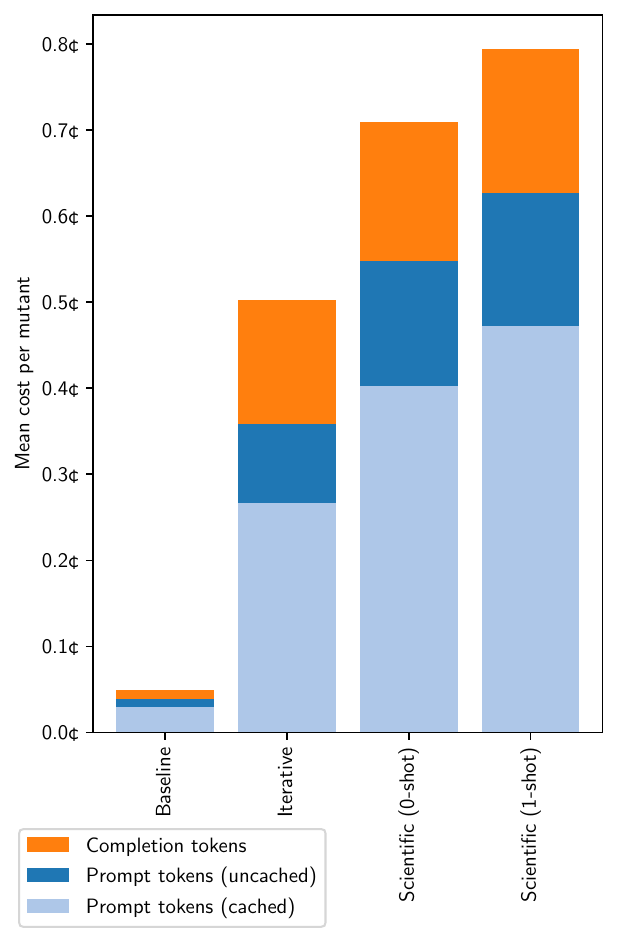}
		\caption{Costs per mutant}
		\label{fig:cost}
	\end{subfigure}
	\hfill
	\begin{subfigure}[t]{0.49\columnwidth}
		\centering
		\includegraphics[width=\textwidth]{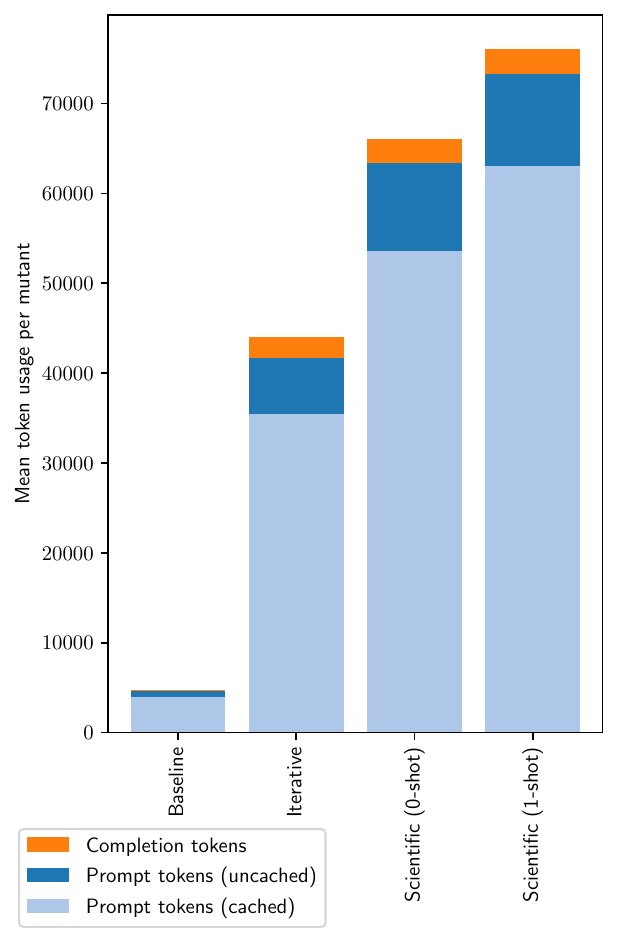}
		\caption{Tokens per mutant}
		\label{fig:token}
	\end{subfigure}

	\caption{Costs and tokens per mutants divided into (1) prompt tokens (cached), (2) prompt tokens (uncached), and (3) completion tokens.}
	\label{fig:coststoken}
\end{figure}

\begin{figure}[t]
	\centering
	\includegraphics[width=\columnwidth]{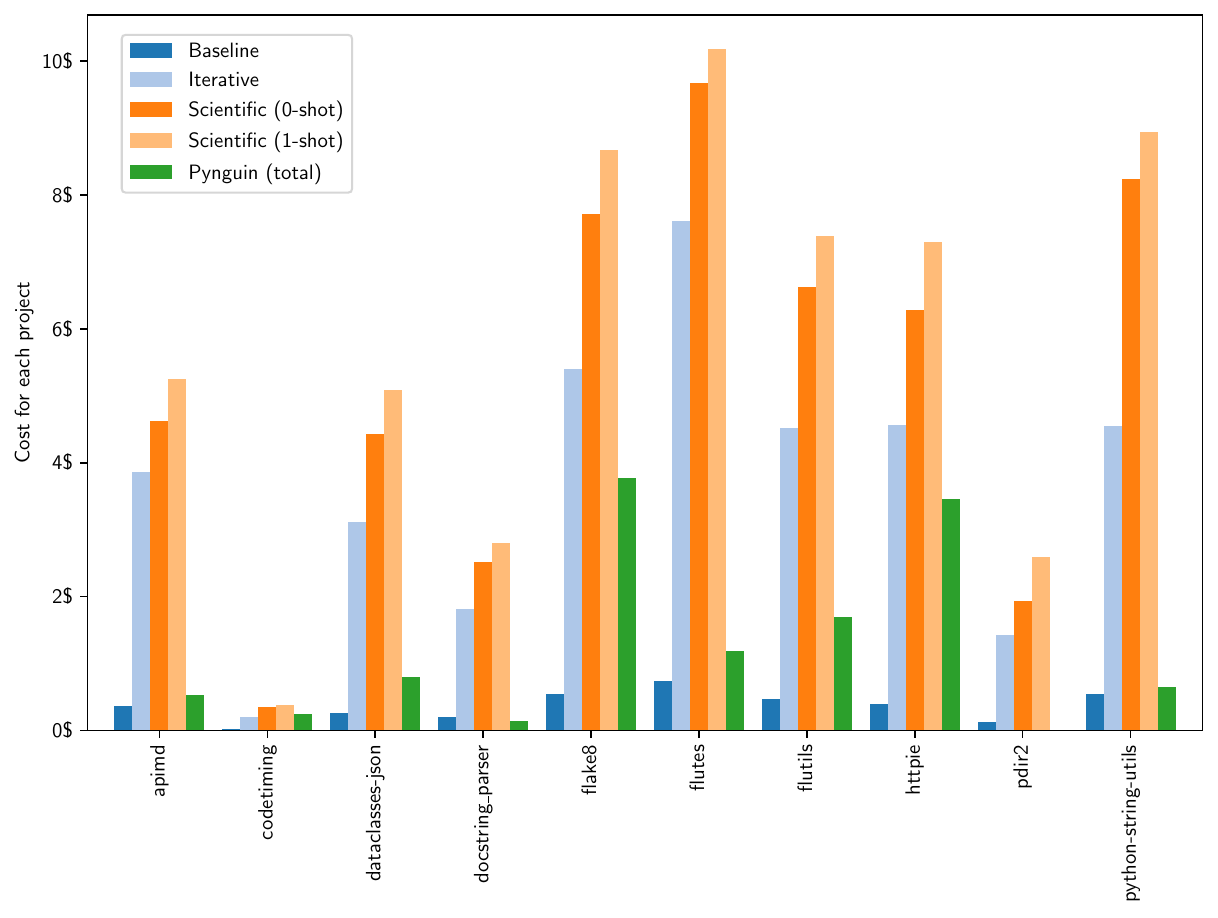}
	\caption{Cost per project and approach.}
	\label{fig:costsprojects}
\end{figure}

\begin{figure}[t]
	\centering
	\begin{subfigure}[t]{0.45\columnwidth}
		\centering
		\includegraphics[width=\textwidth]{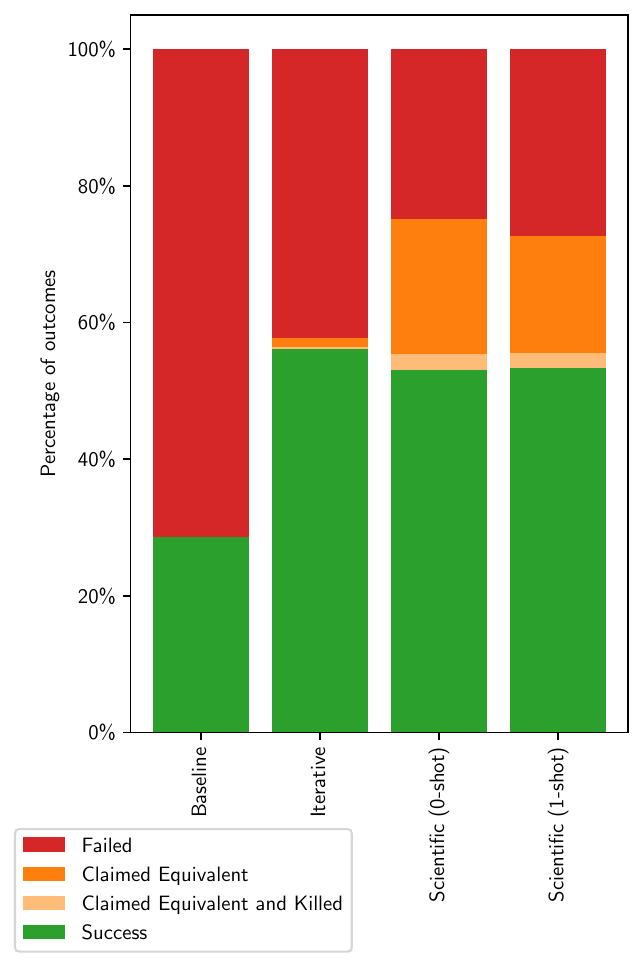}
		\caption{Outcomes (success, claimed equivalent, claimed equivalent and killed, failed).}
		\label{fig:outcome}
	\end{subfigure}
	\hfill
	\begin{subfigure}[t]{0.45\columnwidth}
		\centering
		\includegraphics[width=\textwidth]{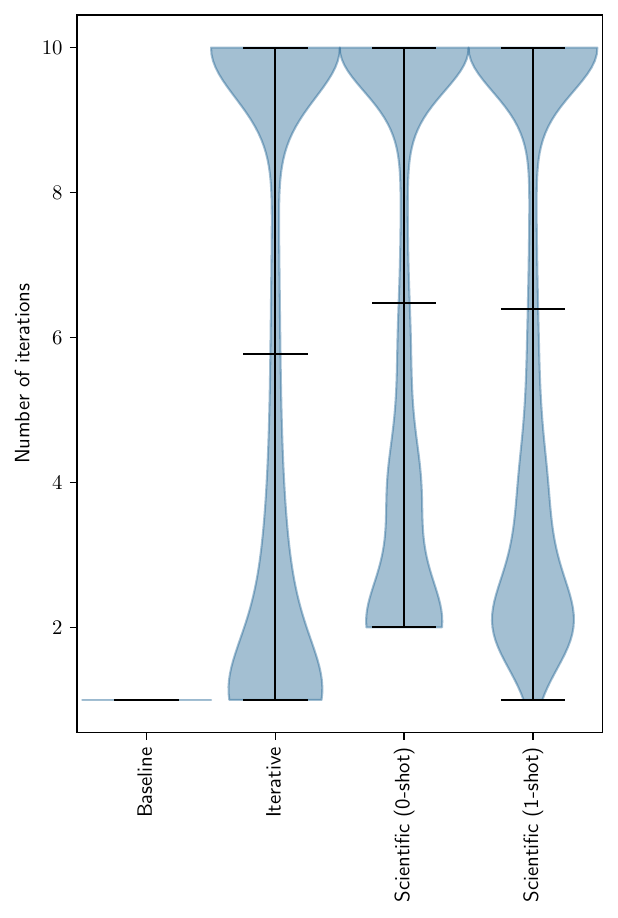}
		\caption{Number of iterations of our approaches.}
		\label{fig:iterations}
	\end{subfigure}

	\caption{Number of iterations and outcomes of our approaches.}
	\label{fig:iterationsoutcome}
\end{figure}

In \cref{fig:cost} and \cref{fig:token}, we show the costs and tokens per mutant. We divide these costs into three categories: cached prompt tokens~(representing prompt prefixes cached by OpenAI), uncached prompt tokens~(all prompt tokens after the first non-cached token), and completion tokens~(representing the LLM's responses). Unsurprisingly, the Baseline was the most cost-effective approach, as it lacks iterations and uses a basic prompt, leaving little opportunity to cache tokens. In contrast, the other approaches could cache most tokens since the input module and mutant remained unchanged during iterations. While completion tokens are few, representing a single test case, they are relatively expensive and contribute significantly to overall costs. Overall, the Baseline incurred the lowest cost, less than \dollar{0.001}, and used around 5,000 tokens per mutant. The other approaches required significantly (\(p<0.001\)) more tokens, with Scientific (1-shot) being the most expensive at approximately \dollar{0.008} and over 70,000 tokens per mutant.

\Cref{fig:costsprojects} reveals significant variations in cost depending on the project. For instance, the project \textit{codetiming} costs less than one dollar across all approaches, whereas the project \textit{flutes} costs approximately ten dollars for each scientific approach, eight dollars for Iterative, and about \dollar{0.80} for the Baseline. This demonstrates that the project size and complexity heavily influence the overall cost, as larger modules increase the prompt size and, consequently, the cost. Compared to the LLM-based approaches, Pynguin is almost always cheaper to run \exnum{30} times than running any of the LLM-based approaches just once.

We present the outcomes for each mutant across all approaches in \cref{fig:outcome}. Mutants fall into three main categories: successfully killed~(a test case was found), failed~(no test case was found), or marked as equivalent~(claimed by the LLM). Since we tell the LLM to continue generating tests after flagging a mutant as equivalent, some mutants were flagged as equivalent and then subsequently killed in the same run. The Baseline achieved significantly~(\(p<0.001\) for all other approaches) the lowest success rate, which aligns with its lack of iterations. The other three approaches had similar success rates of nearly \perc{60}, consistent with findings in \cref{sec:rq1}, where all three iterative methods demonstrated comparable test counts, coverage, and mutation scores. Interestingly, the Baseline did not classify any mutants as equivalent, while Iterative marked a small number as equivalent. Both scientific approaches, however, identified over \perc{15} of mutants as equivalent. We shall explore this further in \cref{sec:rq3}.

\Cref{fig:iterations} illustrates the number of iterations for each approach. As the Baseline does not involve iterations, its value is always one. The other three approaches exhibit similar patterns: most killing tests are found within the first three or four iterations, fewer in the middle stages, and they terminate at ten iterations due to the set maximum. This again suggests that the specific approach used matters less, as long as it is iterative. Given that most tests are identified early, future executions could potentially be limited to fewer than ten iterations.

On average, the two scientific approaches generate a successful test candidate \exnum{1.06} turns later than the iterative approach. While the iterative approach outputs successful test candidates after an average of \exnum{2.51} turns, the scientific approaches take \exnum{3.51} turns (0-shot) and \exnum{3.63} turns (1-shot) to do so. This increase can be explained by our prompt design: The prompt for our scientific approaches instructs the LLM to first write an experiment that can detect the mutant, and only then generate a candidate test, while the iterative approach is instructed to generate test cases right away.

\summary{RQ2}{The Baseline approach was the most cost-effective, requiring fewer tokens and lower costs, while the scientific approaches were significantly more expensive due to the higher number of tokens. Despite the higher cost, the LLM-based approaches showed similar success rates, and the results suggest that using iterations, regardless of the specific approach, tends to yield similar outcomes. Pynguin is cheaper to run \exnum{30} times than a single execution of an LLM-based approach, but results in lower coverage and mutation scores.}

\subsubsection{RQ3: Can LLM-based test generation reliably detect equivalent mutants?}\label{sec:rq3}

\begin{lstlisting}[float, language=diff, float, caption={Equivalence claim with \texttt{len()}.}, captionpos=b, label=lst:01, linewidth=\columnwidth, frame=single, aboveskip=10pt]
-  if received_default and len(unknown) == 0:
+  if received_default and len(unknown) <= 0:
\end{lstlisting}

\begin{lstlisting}[float, language=diff, float, caption={Equivalence claim with regular expression.}, captionpos=b, label=lst:02, linewidth=\columnwidth, frame=single]
   chunk = match.group(0)
   if not chunk:
-    continue
+    break
\end{lstlisting}

\begin{lstlisting}[float, language=diff, float, caption={Equivalence claim with \texttt{split()}.}, captionpos=b, label=lst:03, linewidth=\columnwidth, frame=single, aboveskip=11pt]
-  doc_list.append(doc.split('\n', 1)[0])
+  doc_list.append(doc.split('\n', 2)[0])
\end{lstlisting}

We manually reviewed a random sample of mutants from each project and approach to verify if they were truly equivalent. Out of the 7,380 mutants, 2,427 were flagged as equivalent by at least one of our methods. Of these 2,427 mutants, 1,086 were also killed by one of our approaches when targeted, leaving 1,341 flagged mutants alive. After running all LLM-generated tests against all mutants, \exnum{599} were killed by these tests, leaving \exnum{742} mutants still flagged as equivalent but unkilled. From there, we sampled \exnum{165} mutants for further analysis, including the five mutants per project that were not flagged as equivalent.
For the flagged mutants, we found that in \exnum{98} cases, the mutants could be killed with simple tests, while the remaining \exnum{24} can be considered equivalent.
For the unflagged mutants, we found \exnum{39} mutants killable, and \exnum{4} equivalent.
Projecting from the sample data, we can estimate that only about \(\frac{24}{122} \times 742 \approx 146\) of the 2,427 flagged mutants were actually equivalent.
This is a rather staggering result.
The scientific approaches
flagged substantially more mutants as equivalent (\cref{fig:outcome}),
but the other approaches simply failed to produce tests for the same
mutants, without also flagging them. While scientific
debugging did not support the LLM in detecting equivalent mutants, it
seems that the primary challenge to address first lies in improving
the use of LLMs to identify tests for more challenging mutants.

The \exnum{29} mutants considered equivalent are special
cases worth discussing: They are equivalent when considering only the
PUT and their behavior in tests. However, they could in theory be
killed by leveraging Python-specific language
features~\cite{DBLP:conf/acsc/HolknerH09} (e.g., modifying standard
library methods), as demonstrated with a few examples below.

The mutant in \cref{lst:01} changes \texttt{len(unknown) == 0} to
\texttt{len(unknown) <= 0}. While Python's \texttt{len}
function always returns non-negative integers, a killing test could be
written with a patched \texttt{len} to return negative values, thus
enabling scenarios where mutant and original behave differently.

The mutant in \cref{lst:02} changes \texttt{continue} to \texttt{break} in a block that is logically unreachable because \texttt{group(0)} of a regular-expression match is always present. This means the condition \texttt{if not chunk:} is never satisfied, so the block is unreachable in the original code. A killing test could mock \texttt{re.finditer} to include a match where \texttt{group(0)} returns \texttt{None}, creating a scenario that would trigger the unreachable block. With the mutant applied, this causes the loop to exit prematurely, resulting in incorrect behavior.

The mutant in \cref{lst:03} changes the \texttt{str.split} method call from \verb|split('\n', 1)| to \verb|split('\n', 2)|. However, only the first part of the split string is used in both cases. This change may appear harmless because, in normal circumstances, splitting a string with a limit greater than or equal to 1 would yield the same result for the first part. However, one could write a test that uses a subclass of \texttt{UserString} to mock the \texttt{split} method. The custom \texttt{MockString} asserts that the \texttt{num} parameter passed to \texttt{split} is 1, so it detects the mutant's change when \texttt{num} is 2. This reveals that the mutant is not equivalent, as the mock forces the code to behave differently by catching the modified split behavior.

This means that while the scientific approaches correctly identified such mutants as equivalent when only considering the PUT, one could in theory also classify them as non-equivalent when taking all Python language features into account.

\vspace{0.2cm}
\summarybottom{RQ3}{Many mutants initially flagged as equivalent were later found to be killable, the majority of them being accidentally killable by other LLM-generated tests. Killing a sample of the remaining mutants required only simple tests for most, and advanced Python features for a smaller subset. This highlights the limitations of equivalence detection by using Large Language Models.}

%% file: sections/discussion.tex
The challenges encountered during LLM prompting for generating
experiments and tests highlight several recurring issues. These
challenges provide insight into the limitations of current LLMs in
understanding nuanced task requirements and adhering to specified
formats. Below, we discuss the key problems, their implications, and
the approaches taken to mitigate them.

\paragraph{Assumption of Simultaneous PUT and Mutant Availability}

One prominent issue observed was the LLM's tendency to assume that
both the PUT and Mutant implementations could be invoked from within
the same context, which is often not the case (easily) with mutation
frameworks. This behavior persisted even after explicitly instructing
the LLM on the test execution process--—that tests should pass on the
PUT and fail on the Mutant when run separately. In approximately
one-third of cases, the LLM continued generating tests that relied on
accessing both implementations from within the same context.

Changing the experiment format to explicitly call both implementations did alleviate this behavior but introduced two new problems:

\begin{itemize}
	\item \textbf{Conversion Issues:} The LLM struggled to convert the experiment logic into a valid test format that adhered to the constraint of only accessing one implementation at a time. This limitation resulted in tests that could not be incorporated into the target test suite.
	\item \textbf{Evaluation Challenges:} The previous approach enabled automatic detection of differences between the PUT and Mutant implementations by comparing outputs. The new format, however, complicated this process, as the logic for detecting discrepancies relied on dual invocation, which was incompatible with the single-invocation constraint.
\end{itemize}

\paragraph{Unintended Test Modifications}

Another challenge was the LLM's tendency to modify broken tests by increasing their complexity or length, rather than addressing the core issues. This behavior suggests a potential limitation in the LLM's ability to effectively analyze and manipulate the code it generated. Instead of simplifying or refining tests to meet the required criteria, the LLM produced convoluted outputs that were harder to understand and debug.

\paragraph{Harmful or Erroneous Tests}

Some LLM-generated tests exhibited unintended side effects, such as modifying the code under test or outputting invalid UTF-8 characters that caused crashes. While such behavior may be an expected limitation of LLMs operating in unrestricted programming environments, it underscores the importance of robust safeguards and validation mechanisms in workflows relying on LLMs for automated code generation.

\paragraph{Difficulty in Adhering to Predefined Output Formats}

Achieving consistent adherence to a predefined output structure proved challenging. Despite explicitly listing allowed Markdown headlines and code block formats, the LLM frequently deviated from these instructions, generating unstructured or unexpected outputs. To address this, we implemented lenient matching algorithms to parse and interpret the LLM's responses and employed heuristics to select the most relevant code block from answers containing multiple blocks. While these measures improved reliability, they underscore the difficulty of enforcing strict structural constraints on LLM outputs.

\paragraph{Handling Long Input Files}

The LLM struggled to process long input files, leading to issues when the source code exceeded token limits. The implemented solution—including only the first 100 lines and 900 lines around the mutation—mitigated the issue but was rarely necessary, as test generation focused on projects with shorter source files. Nonetheless, this workaround highlights the need for effective input truncation strategies when dealing with large-scale source code.

%% file: sections/related.tex
In recent years, research has explored generating unit tests using LLMs across various programming languages and models. For instance, studies have used LLMs like ChatGPT to generate tests for JavaScript~\cite{DBLP:journals/tse/SchaferNET24}, and Code Llama for languages such as C\# and C++~\cite{DBLP:conf/fase4games/PaduraruSJ24}.

Much of the recent research has focused on generating unit tests for Java. A notable example is the Integrated Development Environment~(IDE) plugin TestSpark, which generates JUnit tests directly within the IDE~\cite{DBLP:conf/icse/SapozhnikovOPKD24}. TestSpark has been compared to ChatUniTests, a tool that uses ChatGPT, and EvoSuite, a state-of-the-art search-based test generation~(SBTG) tool. In this comparison, ChatUniTests outperformed the other tools in half of the projects evaluated~\cite{DBLP:journals/corr/abs-2305-04764}. Other studies have also compared ChatGPT-based test generation with EvoSuite, reporting that even basic prompts can be as effective, if not more so, than current state-of-the-art methods~\cite{DBLP:journals/tse/TangLZL24,DBLP:conf/sast/GuilhermeV23,DBLP:journals/corr/abs-2406-18181}. The key difference in our approach is that while these tools rely on simple prompts, we employ a more sophisticated approach using scientific debugging.

Other recent efforts have aimed to improve test generation with LLMs in various ways. CodaMosa~\cite{DBLP:conf/icse/LemieuxILS23} builds on Pynguin and queries an LLM for a new test case if Pynguin's search is stuck; it adds the test case from the LLM to Pynguin's population and can achieve higher coverage results. Other combinations of Pynguin and an LLM to enhance the test generation process also showed positive results across all steps~\cite{DBLP:conf/internetware/XiaoGLC24}. In contrast, our method relies solely on the LLM without integrating any existing SBTG tools. Another example is TestART which first generates an initial test suite and then asks the LLM to correct failing tests~\cite{DBLP:journals/corr/abs-2408-03095}. Similarly, ChatTester improves the quality of generated tests in a second step, though it does not fix failing tests~\cite{DBLP:journals/pacmse/Yuan0DW00L24}. All approaches use iterative improvements, but our tool directly implements an iterative process driven by the LLM, without external intervention. A recent survey analyses more than 100 relevant studies using LLMs for software testing~\cite{DBLP:journals/tse/WangHCLWW24}, which underlines the relevance of the topic.

The tool most similar to ours is MuTAP~\cite{DBLP:journals/infsof/DakhelNMKD24}, which first generates a test suite and then refines it by having the LLM generate new tests based on surviving mutants. This is also an iterative approach, but it operates from the outside. In contrast, our method enables the LLM to iteratively refine the tests within a single prompt, guided by both scientific debugging and mutation testing. Since MuTAP relies on an external iterative process, while our approach integrates refinement directly into one interaction, the two methods differ fundamentally. As such, a direct comparison would not yield meaningful insights into their respective strengths.

%% file: sections/conclusions.tex
In this paper, we explored the potential of large language models~(LLMs) for automated test generation, focusing on their ability to create high-quality test cases and detect software faults in Python projects. Using mutation analysis as a benchmark, we systematically evaluated the performance of LLM-generated tests across diverse metrics, including code coverage, mutation scores, and the detection of equivalent mutants. Our empirical study provides valuable insights into the strengths and limitations of LLMs in this domain. We found that LLMs demonstrate promising capabilities in generating test cases that achieve competitive code coverage and mutation scores. However, their performance varies significantly across projects, emphasizing the importance of understanding the factors influencing their effectiveness. Additionally, our findings highlighted the critical role of prompt design in determining the quality of the generated tests, suggesting opportunities for further optimization.

A key takeaway from our work is that the specific LLM-based approach used for automated test generation matters less than ensuring that the process is iterative. Iterative refinement, whether in the form of optimizing prompts for LLMs or improving test strategies in traditional tools, consistently leads to better results. Our study also revealed that while the costs associated with LLM-based approaches are significantly higher than those for traditional tools like Pynguin, the improved quality of the results may justify these costs. LLM-generated tests showed better fault detection and coverage, making them a valuable option despite the increased computational expense.

While our study highlights the potential of LLMs, it further identifies challenges that need to be addressed to fully integrate these models into automated testing workflows. Future work shall extend the dataset to include projects from other programming languages and domains, thereby evaluating the generalisability of LLMs in diverse software ecosystems. Furthermore, developing methods to automate and optimize prompt design could enhance the quality and reliability of LLM-generated test cases. Another promising direction is the investigation of hybrid approaches that integrate LLM-based test generation with traditional methods, such as symbolic execution, to achieve improved results. Finally, exploring the effective adoption of LLMs in industrial settings, including addressing challenges such as scalability and integration with existing software development pipelines, represents an important step forward.